\documentclass[10pt,a4paper]{article}
\usepackage[latin1]{inputenc}
\usepackage{amsmath}
\usepackage{amsfonts}
\usepackage{amssymb}
\usepackage{graphicx}
\begin{document}
\large
\textbf{ Observational Constraints on Model with specific q}\\
\begin{center}
A. Beesham\footnote{Email: abeesham@yahoo.com} \\
\vspace{2mm}
 \small 
 Faculty of Applied and Health Sciences, Mangosuthu University of Technology, Jacobs 4026,   South Africa\\
 
\end{center}
\vspace{3mm}
\textbf{Abstract:} We place observational constraints on an FLRW cosmological model in $f(R,L_m)$ gravity with a specific deceleration parameter that depends on the scale factor. This form of the deceleration parameter  has been discussed by   authors in several papers, but none of them have applied observations to constrain the variables of the model. We carry this out with the cosmic chronometer, supernovae and the baryon acoustic oscillation datasets. The optimum values for the relevant parameters are found and used to plot the kinematical and physical parameters of the model.  Although  the model tends to the standard Lambda cold dark matter model at late times,  there are several  issues with the model concerning the values of some of the parameters and the energy conditions. The transition redshift of the model does not match with Planck data. The equation of state parameter indicates that the model falls into the category of phantom dark energy, which is not well supported by observations. Thus the model does not seem viable.
	
\section{Introduction:} The deceleration parameter $q$ in cosmology is defined as:
\begin{equation}\label{defnqa}
	q = -\frac{\ddot{a}a}{\dot{a}^2}   
\end{equation}
where $a$ is the scale factor, and a dot denotes a time derivative.
Observations indicate that the universe was decelerating in the past, but that, as from a certain time onwards, the universe started accelerating  \cite{Riess,Perl}. Cosmology has to be  able to explain this. The standard $\Lambda$CDM model in general relativity has the most support amongst the community to explain the current acceleration. In this model, the existence of some hypothetical matter with negative pressure is postulated. This exotic matter causes the present acceleration.  However, the model has several  shortcomings \cite{Durrer,Abd}, such as  the cosmic coincidence and cosmological constant problems.  Hence, there is a strong motivation to study modified gravity theories  with a view towards explaining the  transition, as well as solving some of the problems of the $\Lambda$CDM model. Many modified theories possess solutions with late time acceleration without matter that has negative pressure. However, there are situations when,   to get solutions,  researchers in the field have to make a choice for some parameter, such as the deceleration parameter, Hubble parameter, scale factor,   or equation of state parameter. 

In this work, we investigate cosmological models in  $f(R,L_m)$ modified gravity, where $L_m$ is the matter Lagrangian, with $R$ being the Ricci scalar.  A specific choice of the form of the deceleration parameter is considered, viz., 
\begin{equation}
	q = -1+\frac{\eta}{1+ a^\eta}. \label{qformeq}
\end{equation}
where $\eta > 0$ is a positive constant. One very obvious motivation for this is that at late times, for large scale factor $a$ (assuming that for the model, the scale factor $a$ is increasing at late times. This may not always be the case, e.g., for oscillating models), the deceleration parameter $q$ tends to $-1$, the value for the  $\Lambda$CDM model at late times. Hence the assumption of this condition ensures that the model will approach the $\Lambda$CDM model in future, or equivalently, for large $a$. Another motivation is that, as we shall see later in detail,  it has only two parameters that need to be constrained by observations. We shall discuss this further at a later stage.

We first describe briefly the models in the literature with this form of the deceleration parameter $q$.
Singha and Debnath \cite{Singha}  investigated a quintessential cosmological model, with a minimally
coupled scalar field, assuming the form (\ref{qformeq}) of the deceleration parameter $q$. A barotropic fluid
and models dominated by Chaplygin gas provide a transition from  deceleration to  acceleration. For this, the potential function $V (\phi)$ is always decreasing with the scalar field $\phi$.  The behaviour of the models at  different stages during their evolution was illustrated by plots of the statefinder parameters $(r, s)$ (these parameters will be defined later).

Chirde and Shekh \cite{CS} investigated a symmetric  plane non-static   space-time in $f (R, T)$  gravity containing a perfect fluid, with $T$ being the trace of the energy-momentum tensor $T_{ab}$.  In order to get a specific solution, a special form of $q$ as in (\ref{qformeq}) was used. The other relation that was assumed was that the equation of state parameter and the metric potentials  are proportional to the skewness parameter. The anisotropy initially increases,  reaching a maximum at some finite time, and then decreases to zero in keeping with current observations. For $n \geq 1$, the model initially decelerates, and then later accelerates.  Some kinematical and physical properties of the model was also studied.

The study of the spatially homogeneous and anisotropic Bianchi type-I universe in $f(R, T)$ gravity was undertaken by Sahoo et al \cite{Sahoo}. They assumed two different forms of  $f(R, T)$,   viz., $f(R, T) = f_2(T)+f_1(R)$ and $f(R, T) = 2f(T)+R $ and
and incorporated bulk viscosity.  A form (\ref{qformeq}) of the  deceleration parameter   was employed. Exact solutions of the field equations were found, and the kinematical and physical   properties of both sets of  models were  studied in detail as far as the future  of the universe as concerned.  The nature of the weak, dominant and strong energy conditions for both cases were investigated. The analysis revealed that both models incorporating a bulk viscosity matter component exhibit an acceleration of the universe during late times. Furthermore, the cosmic jerk parameter aligns well with the three kinematical data sets. All energy conditions were also investigated.

Recently, Pawde et al. [8] conducted an investigation into an anisotropic model within the framework of $f(R,L_m)$ gravity. They adopted the specific form $f(R,L_m) = R^2 + L^{\alpha}_m + \beta$, where $\beta$ represents a constant. This study focused on elucidating the dynamics of the universe by examining a variable deceleration parameter of the type (2). Furthermore, the researchers employed energy conditions, jerk, equation of state, and state-finder parameters to enhance their understanding of the universe's evolution in this modified gravity context. They reported that their results were consistent with recent observational data and aligned with the $\Lambda$CDM model. Additionally, they asserted that their research offers significant insights into the anisotropic characteristics of the universe as described by $f(R,L_m)$ gravity, demonstrating its proximity to the $\Lambda$CDM model and thereby enriching the comprehension of the fundamental dynamics governing cosmic evolution. However, it is noteworthy that no observational diagrams depicting cosmic chronometer, supernovae or baryon acoustic  background data were included, nor was any MCMC analysis presented.

In this investigation, we constrain the variables of the model studied in ref. \cite{Pawde} by utilising observations. We carry this out with the cosmic chronometer, supernovae and baryon acoustic  datasets. The optimum values for the relevant variables are found, as opposed to the ad hoc choices that Pawde et al. \cite{Pawde} made. We plot the kinematic parameters of the model, such as the state-finder parameters $(r,s)$,   jerk parameter $j$,  and deceleration parameter $q$. It is found that our values lead to the model tending to the standard $\Lambda$CDM model in future, and exhibiting a transition from deceleration to acceleration. As in ref \cite{Pawde}, we work  in $f(R,L_m)$ gravity, and we also study the physical parameters such as the energy, pressure, equation of state parameter and the   energy conditions.
 
 \section{Basic Formalism of $f(R, \mathcal{L}_m)$ Gravity Theory}
	
	With the Ricci scalar \( R \) and the matter Lagrangian density \( \mathcal{L}_m \),  the action principle for the \( f(R, \mathcal{L}_m) \) gravity model introduced by Harko and Lobo (2010) \cite{28} is expressed as:
	\begin{equation}\label{1}
		S = \int f(R, \mathcal{L}_m) \sqrt{-g} \, d^4x,
	\end{equation}
	where $f$ is an arbitrary function of $R$ and $\mathcal{L}_m$.
	The Ricci scalar \(R\) is defined using the Ricci tensor and the metric tensor \(g^{ij}\):
	\begin{equation}
		R = g^{ij} R_{ij},
	\end{equation}
	where the Ricci tensor is given by
	\begin{equation}
		R_{ij} = \partial_\kappa \Gamma^\kappa_{ij} - \partial_j \Gamma^\kappa_{i\kappa} 
		+ \Gamma^\lambda_{ij} \Gamma^\kappa_{\lambda\kappa} 
		- \Gamma^\kappa_{j\lambda} \Gamma^\lambda_{i\kappa}.
	\end{equation}
In this context, $\Gamma^i_{j\kappa}$ denotes the components of the Levi-Civita connection, which is defined as:
\begin{equation}
	\Gamma^i_{j\kappa} = \frac{1}{2} g^{i\lambda} \left( 
	\frac{\partial g_{\lambda\kappa}}{\partial x^j} + 
	\frac{\partial g_{\lambda j}}{\partial x^\kappa} - 
	\frac{\partial g_{j\kappa}}{\partial x^\lambda}
	\right).
\end{equation}

The field equations for \( f(R, \mathcal{L}_m) \) gravity, obtained by varying the action principle \eqref{1} with respect to the metric tensor \( g_{ij} \), is given by:
\begin{equation}
	f_R R_{ij} - \frac{1}{2} f g_{ij} + \left( g_{ij} \Box - \nabla_i \nabla_j \right) f_R 
	= \frac{1}{2} f_{\mathcal{L}_m} T_{ij},
\end{equation}
where
\begin{equation}\label{5}
	f_R(R, \mathcal{L}_m) = \frac{\delta f(R, \mathcal{L}_m)}{\delta R}, \quad 
	f_{\mathcal{L}_m}(R, \mathcal{L}_m) = \frac{\delta f(R, \mathcal{L}_m)}{\delta \mathcal{L}_m},
\end{equation}
The symbol $\Box = \nabla^i \nabla_i$, and $T_{ij}$ is the energy-momentum tensor (EMT) for a perfect fluid, given by
\begin{equation}
	T_{ij} = -\frac{2}{\sqrt{-g}} \frac{\delta \left(\sqrt{-g} \mathcal{L}_m\right)}{\delta g^{ij}} = \mathcal{L}_m g_{ij} - 2 \frac{\partial \mathcal{L}_m}{\partial g^{ij}}.
\end{equation}
From the explicit form of the field equations, we can determine the covariant divergence of the energy-momentum tensor \( T_{ij} \) as:
\begin{equation}
	\nabla^i T_{ij} = 2 \nabla^i \ln(f_{\mathcal{L}_m}) \frac{\partial \mathcal{L}_m}{\partial g^{ij}}.
\end{equation}
Additionally, by contracting the field equations, we can derive the relationship between the Ricci scalar \( R \), the matter Lagrangian density \( \mathcal{L}_m \), and the trace \( T \) of the stress-energy-momentum tensor \( T_{ij} \):
\begin{equation}
	R f_R + 3 \Box f_R - \left( f - f_{\mathcal{L}_m} \mathcal{L}_m \right) = \frac{1}{2} f_{\mathcal{L}_m} T,
\end{equation}
where, for any function $F$, 
\begin{equation}
	\Box F = \frac{1}{\sqrt{-g}} \partial_i \left( \sqrt{-g} g^{ij} \partial_j F \right).
\end{equation}
\section{Metric and equations of motion in $f(R, \mathcal{L}_m)$ Gravity }
It is worth noting that anisotropic models add extra complexity and degrees of freedom to the equations governing the dynamics of the universe, making them more difficult to handle and validate against observational data than the simpler isotropic models, such as the FLRW (Friedmann-Lemaitre-Robertson-Walker) models.
Therefore, to examine the anisotropic features of the universe, we shift our focus to studying the anisotropic and spatially homogeneous locally rotationally symmetric (LRS) Bianchi type-I metric given by
\begin{equation}\label{9}
	ds^2 = -dt^2 + A^2(t) dx^2 + B^2(t) \left( dy^2 + dz^2 \right),
\end{equation}
where \( A(t) \) and \( B(t) \) depend only on the cosmic time \( t \). When \( A(t) = B(t) = a(t) \), the FLRW cosmological model is recovered. The adoption of the LRS Bianchi I model also facilitates comparison with that of ref  \cite{Pawde}. 
The Ricci scalar associated with the LRS Bianchi-I spacetime can be expressed as:
\begin{equation}
	R = -2 \frac{\ddot{A}}{A} + 2 \frac{\ddot{B}}{B} + 2 \frac{\dot{A} \dot{B}}{A B} + 2 \frac{\dot{B}^2}{B^2},
\end{equation}
where the dots denote derivatives with respect to cosmic time \( t \).

Here, we consider that the matter distribution of the universe is described by a perfect fluid with  energy-momentum tensor (EMT):  
\begin{equation}
	T^i_j = (p + \rho) u^i u_j + p g^i_j,
\end{equation}
where \( p \) and \( \rho \) represent the thermodynamic pressure and energy density of the matter, respectively, and \( u^i = (1, 0, 0, 0) \) are the components of the comoving four-velocity vector in the cosmic fluid, satisfying \( u^i u_j = 0 \) and \( u^i u_i = -1 \).
The field equations \eqref{5}, which govern the dynamics of the Universe in \( f(R, \mathcal{L}_m) \) gravity, are expressed as:
\begin{equation}\label{12}
	-\frac{\ddot{A}}{A} + 2 \frac{\dot{A} \dot{B}}{AB} f_R 
	- \frac{1}{2} \left( f - f_{\mathcal{L}_m} \mathcal{L}_m \right) 
	- 2 \frac{\dot{B}}{B} \dot{f}_R - \ddot{f}_R 
	= \frac{1}{2} f_{\mathcal{L}_m} p,
\end{equation}
\begin{equation}\label{13}
	-\frac{\ddot{B}}{B} + \frac{\dot{B}^2}{B^2} + \frac{\dot{A} \dot{B}}{AB} f_R 
	- \frac{1}{2} \left( f - f_{\mathcal{L}_m} \mathcal{L}_m \right) 
	- \left(\frac{\dot{A}}{A} + \frac{\dot{B}}{B}\right) \dot{f}_R - \ddot{f}_R 
	= \frac{1}{2} f_{\mathcal{L}_m} p,
\end{equation}
\begin{equation}\label{14}
	-\frac{\ddot{A}}{A} + 2 \frac{\ddot{B}}{B} f_R 
	- \frac{1}{2} \left( f - f_{\mathcal{L}_m} \mathcal{L}_m \right) 
	- \left(\frac{\dot{A}}{A} + \frac{\dot{B}}{B}\right) \dot{f}_R 
	= -\frac{1}{2} f_{\mathcal{L}_m} \rho,
\end{equation}
where dots represent derivatives with respect to cosmic time $t$, and $f_R = \frac{\partial f}{\partial R}$ and $f_{\mathcal{L}_m} = \frac{\partial f}{\partial \mathcal{L}_m}$.

\section{Cosmological solutions for $f(R, \mathcal{L}_m)$ gravity }
To analyze the dynamics of the Universe, we use the functional form of \( f(R, \mathcal{L}_m) \) gravity as:  
\begin{equation}
	f(R, \mathcal{L}_m) = \frac{R}{2} + \mathcal{L}_m^\alpha + \beta,
\end{equation}
where \( \alpha \) and \( \beta \) are arbitrary parameters, and the standard field equations of general relativity (GR) are recovered for \( \alpha = 1 \).
For this specific functional form of \( f(R, \mathcal{L}_m) \) gravity, we consider \( \mathcal{L}_m = \rho \) \cite{14,46,47,48,49}, so that the field equations \eqref{12}, \eqref{13}, and \eqref{14} give:
\begin{equation}\label{16}
	2 \frac{\ddot{B}}{B} + \frac{\dot{B}^2}{B^2} - \beta - (1 - \alpha) \rho^\alpha = \alpha \rho^{\alpha - 1} p,
\end{equation}
\begin{equation}\label{17}
	\frac{\ddot{A}}{A} + \frac{\ddot{B}}{B} + \frac{\dot{A} \dot{B}}{AB} - \beta - (1 - \alpha) \rho^\alpha = \alpha \rho^{\alpha - 1} p,
\end{equation}
\begin{equation}\label{18}
	\frac{\dot{B}^2}{B^2} + 2 \frac{\dot{A} \dot{B}}{AB} - \beta = (1 - 2\alpha) \rho^\alpha.
\end{equation}
The anisotropic parameter $\Delta$ is given by  
\begin{equation}
	\Delta = \frac{1}{3} \sum_{i=1}^3 \frac{H_i}{H} - 1,
\end{equation} and in this case, works out to be:  
\begin{equation}
	\Delta = \frac{1}{2}.
\end{equation}
The shear scalar is given by  
\begin{equation}
	\sigma^2 = \frac{3}{2} \Delta H^2,
\end{equation}
We encounter a system of three equations, given in Eqs. (\ref{16})-(\ref{18}), which involve four unknowns: \(A\), \(B\), \(p\), and \(\rho\). Therefore, to obtain a complete solution, it is essential to introduce an additional reasonable condition to determine a unique and deterministic solution for this system.

\section{Solving field equations via model independent technique}

We first discuss the kinematical quantities associated with the choice of function for $q$ as in equation (\ref{qformeq}).   The Hubble parameter $H(a)$ is defined as $H(a) \equiv \dot{a}/a$. From equation (\ref{qformeq}), the Hubble parameter can be obtained as: 
\begin{equation}\label{Haeq}
	H(a) \equiv \frac{\dot{a}}{a} = \xi (1+a^{-n})
\end{equation}
where $\xi$ is an  integration constant.
The scale factor  $a(t)$ as a function of time $t$ can be calculated from $H=\dot{a}/a$:
\begin{equation}\label{ateq}
	a(t) = (e^{\eta t}-1)^{1/ \eta}
\end{equation}
Now the redshift $z$ in terms of $a$ is given by:
\begin{equation}\label{azdefn}
	a=(1+z)^{-1} 
\end{equation}
where $z$ is the redshift. Hence, we find the Hubble parameter in terms of redshift from equations (\ref{qformeq}), (\ref{Haeq})-(\ref{azdefn}) as:
\begin{equation}\label{Hzformeq}
	H=\frac{H_0}{2}[1+  (1+z)^{\eta}].
\end{equation}
where $H_0$ is the current value of the Hubble parameter. The deceleration parameter in terms of redshift from equations (\ref{qformeq}) and (\ref{azdefn}) is given by:
\begin{equation}\label{qzformeq}
	q(z) = q = -1+\frac{\eta}{1+ (1+z)^{- \eta}}
\end{equation}

We now  provide additional  motivation for our choice of deceleration parameter $q$ as in Eq. (\ref{qzformeq}). The study of cosmological models within the climate of  late time acceleration relies heavily on kinematic variables. The deceleration parameter, e.g., characterizes the behavior of the universe, such as whether it is undergoing deceleration, acceleration, or in a transition phase. The EoS parameter $\omega$ provides insight into the physical properties of the energy sources driving the evolution of the universe. In addition, to determine other cosmological parameters, such as pressure, energy density, EoS parameter, and potential function, an extra equation is usually needed needed to complete the system of field equations, i.e., Eqs. (\ref{16})-(\ref{18}). This 	supplementary equation can be any functional form of a cosmological parameter, such as the Hubble parameter, deceleration parameter, or EoS parameter, providing the necessary constraint equation/s \cite{Pacif}.
	
The motivation behind a  parametrization of the deceleration parameter  comes from the fact that the
deceleration parameter is an important cosmological parameter that characterizes the dynamics of the universe. It is defined as the ratio of the cosmic acceleration to the cosmic expansion rate, i.e., $q = -\frac{\ddot{a}}{aH^2}$. In the past, the deceleration parameter was thought to be constant, indicating that the universe is either slowing down or maintaining a constant rate of expansion. However, observational evidence in recent years suggests that the universe is in fact accelerating at present,  which requires a modification of the deceleration parameter. To account for this acceleration, various parametrizations of the deceleration parameter have been proposed in the literature \cite{Kous} (and references therein). For our choice of $q$, we can make some qualitative remarks about how $q$ varies with redshift. 
\begin{itemize}
	\item The parameter $\eta$ relates to the present value of the deceleration parameter, i.e., $q_0 = -1 + \eta/2$. The present state of the universe depends upon the value of $\eta$. If $\eta = 2$, then $q_0 = 0$, and the universe if undergoing constant expansion. If $\eta > 2$, then $q_0 > 0$, and we have decelerated expansion. Finally, if $\eta < 0$, then $q_0 < 0$, and we have accelerated expansion. 
	\item In the distant past, $z>>1$, and $q(z) \rightarrow -1 + \eta$. For $\eta > 1$, we have $q>0 \implies$ deceleration, i.e., the radiation and matter dominated eras.
	\item In the far future, from the form of $q(a)$ as in Eq. (\ref{qformeq}), and the discussion following that equation, we have seen that $q  \rightarrow -1$, the asymptotic form for $q$ for the $\Lambda$CDM model. Hence this model will asymptotically approach the $\Lambda$CDM model in future.
\end{itemize} 

From the definition of spatial volume in terms of the average scale factor, i.e., $V = a^3 = AB^2$, and using Eqs. \eqref{16}-\eqref{18} and Eq. \eqref{ateq}, the corresponding metric potentials $A$ and $B$ are obtained as:
	\begin{equation}\label{28}
		A = \left( e^{\eta t} - 1 \right)^{\frac{2}{\eta}}, \quad B = \left( e^{\eta t} -1 \right)^{\frac{1}{2\eta}}.
	\end{equation}
	The shear scalar $\sigma$ is given by
	\begin{equation}
		\sigma^2 = \frac{3}{4} \left( \frac{e^{\eta t}}{e^{\eta t} - 1} \right)^2.
	\end{equation}
	Corresponding to the above metric potentials, the line element in Eq.~\eqref{9} becomes:
	\begin{equation}
		ds^2 = -dt^2 + \left( e^{\eta t} - 1 \right)^{\frac{4}{\eta}} dx^2 + \left( e^{\eta t} - 1 \right)^{\frac{1}{2\eta}} \left( dy^2 + dz^2 \right).
	\end{equation}
	Solving Eqs. (\ref{16}) and (\ref{18}), and using Eq. \eqref{28}, the energy density $\rho$ in terms of $t$ and $z$ is given by:
  \begin{equation}\label{rhot}
 	\rho(t) = 
 	\left[ \frac{1}{1 - 2\alpha}
 	\left( 
 	\frac{9}{4} \frac{e^{2\eta t}}{(e^{\eta t} - 1)^2} 
 	- \beta 
 	\right)\right]^{\frac{1}{\alpha}}
 \end{equation}
 \begin{equation}\label{rhoz}
 	\rho(z)=\left[ \frac{9\left( \frac{1}{1+z}\right)^{-2\eta}\left( \left( \frac{1}{1+z}\right)^{\eta}+1\right)^2}{4(1-2\alpha)}-\beta\right]^{\frac{1}{\alpha}}
 \end{equation}
 The pressure $p$ in terms of $t$ and $z$ is
 \begin{equation}\label{prest}
 	p(t) = \frac{ -\frac{9\alpha}{4(1 - 2\alpha)} \frac{e^{2\eta t}}{(e^{\eta t} - 1)^2} 
 		- \frac{4\eta e^{\eta t}}{(e^{\eta t} - 1)^2} 
 		+ \frac{\alpha \beta}{1 - 2\alpha}}{\alpha\left[
 		\frac{1}{1 - 2\alpha} 
 		\left( 
 		\frac{9}{4} \frac{e^{2\eta t}}{(e^{\eta t} - 1)^2} - \beta 
 		\right)
 		\right]^{\frac{\alpha - 1}{\alpha}}}  
 \end{equation}
 \begin{equation}\label{presz}
 	p(z) =\frac{\left[
 		\frac{\alpha \beta}{1 - 2\alpha} 
 		-  \frac{9\alpha\left( \frac{1}{1+z}\right)^{-2\eta}\left( \left( \frac{1}{1+z}\right)^{\eta}+1\right)^2}{4(1-2\alpha)}
 		- 4\eta\left( \frac{1}{1+z}\right)^{-2\eta}\left( \left( \frac{1}{1+z}\right)^{\eta}+1\right)
 		\right]}{\alpha\left[ \frac{9\left( \frac{1}{1+z}\right)^{-2\eta}\left( \left( \frac{1}{1+z}\right)^{\eta}+1\right)^2}{4(1-2\alpha)}-\beta\right]^{\frac{\alpha -1}{\alpha}}} 
 \end{equation}
The equation of state parameter $\omega$ in terms of $t$ and $z$ is given by 
\begin{equation}
	\omega(t) =\frac{p}{\rho} =\frac{ -\frac{9\alpha}{4(1 - 2\alpha)} \frac{e^{2\eta t}}{(e^{\eta t} - 1)^2} 
		- \frac{4\eta e^{\eta t}}{(e^{\eta t} - 1)^2} 
		+ \frac{\alpha \beta}{1 - 2\alpha}}{\frac{\alpha}{1-2\alpha}\left[
		\frac{1}{1 - 2\alpha} 
		\left( 
		\frac{9}{4} \frac{e^{2\eta t}}{(e^{\eta t} - 1)^2} - \beta 
		\right)
		\right]}
\end{equation}
\begin{equation}\label{eosz}
	\omega(z) = \frac{\left[
		\frac{\alpha \beta}{1 - 2\alpha} 
		-  \frac{9\alpha\left( \frac{1}{1+z}\right)^{-2\eta}\left( \left( \frac{1}{1+z}\right)^{\eta}+1\right)^2}{4(1-2\alpha)}
		- 4\eta\left( \frac{1}{1+z}\right)^{-2\eta}\left( \left( \frac{1}{1+z}\right)^{\eta}+1\right)
		\right]}{\alpha\left[ \frac{9\left( \frac{1}{1+z}\right)^{-2\eta}\left( \left( \frac{1}{1+z}\right)^{\eta}+1\right)^2}{4(1-2\alpha)}-\beta\right]}
\end{equation}
\section{Observational constraints}
In this section, a statistical analysis is performed to compare the predictions of the theoretical model with observational data. The goal is to establish constraints on the free parameters of the model, namely $H_0$ and $\eta$.  

The CC (cosmic chronometer)  , SNIa (Pantheon), and the BAO datasets are used. The first two datasets are made up of  31 and 1048 data points, respectively.   A software package in Python \cite{DFM} is used, implementing Markov Chain Monte Carlo (MCMC) analysis which implements Bayesian methods in cosmology \cite{Hob}. To estimate the posterior distribution of the model parameters, the MCMC sampler is used. Multiple iterations of MCMC sampling are used to compute the posterior distribution, and observational data is used to construct the likelihood function. To get the best-fitting results, we use 1000 steps and 100 walkers in our MCMC analysis.  The likelihood function is \cite{Myrza,Erre}:
\begin{equation}
	L \propto  exp(-\chi^2/2)
\end{equation}
where the pseudo chi-squared function is given by $\chi^2$ \cite{Hob}.  This seems to be the most common likelihood function $L$ used by observational cosmologists, and hence we have adopted this form. The motivation for this particular form is the following. The formula for the probability density function for a normal distribution contains the term . So when one takes the log of that function, it becomes just the usual $-x^2/2$. Now in basic MCMC analysis, we want to calculate the posterior probability function, which is nothing but usually a joint probability function for a normal distribution case. So in this case, the posterior probability for the joint case becomes $exp[-\sum(-x^2/2)]$. Now we have to find the maximum of this posterior probability. So for ease of calculation, we calculate the maximum of log(Posterior), which simply becomes a $(-x^2/2)$ function. Now these two optimization problems, viz., firstly to calculate the maximum of the log(Posterior), and secondly, the minimum of of $(-x^2/2)$ become equivalent. So the expression $exp(-x^2/2)$ is due to the normal distribution, and we take the log of the Likelihood to make thins simpler.  Given a particular L, it is sometimes easier to rewrite the likelihood as 
	the log of the likelihood. This entails no loss of generality as maximizing a log likelihood 
	is the same as maximizing a likelihood.  There are several other likelihood functions and methods that can potentially be used, but many of them are still in the trial/developmental phase. Some of these are machine learning emulators \cite{Gong}, partition function approach \cite{Rover}, renormalisation group computation \cite{Donald}, principal component analysis \cite{Lin}, Copula method (should yield a more accurate likelihood function in future \cite{Sato1,Sato2}.
	
\subsection{Observational $H(z)$ data}

By measuring the relative ages of galaxies that are passively evolving, red{the} CC dataset can estimate the Hubble parameter. Galaxies can be recognized by certain characteristics in their color profiles and spectra. \cite{R35}. 
CC get their data from the estimated ages of galaxies at different redshifts. Use is made of the 31 independent \( H(z) \) observations in the redshift range \( 0.07 \leq z \leq 2.41 \) in the   analysis \cite{R36}. The formula $ H(z) = -\frac{1}{1+z} \frac{dz}{dt}$ is used to acquire these \( H(z) \) measurements. In this case, \( \frac{dz}{dt} \) is taken as \( \frac{\Delta z}{\Delta t} \), where \( \Delta z \) and \( \Delta t \) indicate the different redshifts and corresponding ages of two galaxies. Then the  corresponding \(\chi^2\) function is:
\begin{equation}\label{4a}
	\chi_{Hz}^{2}=\sum\limits_{k=1}^{31}\frac{[H_{th}(z_{i})-H_{obs,i}]^{2}}{\sigma _{i}^{2}}.  
\end{equation}
where
	\begin{itemize}
		\item \(\mu_{\text{obs},i}\) is the observed distance modulus of the \(i\)-th object,
		\item \(\mu_{\text{th},i}\) is the theoretical distance modulus predicted by the cosmological model,
		\item \(\sigma_i\) is the uncertainty in the observed distance modulus.
	\end{itemize}
	The best-fit values of the model parameters are obtained by minimizing the chi-square statistic \(\chi^2\). This is illustrated in  Figure 1 corresponding to the  1$\sigma$ and 2$\sigma$ error bars obtained from the $CC$ data. 
The best-fit values of the model parameters are obtained by minimizing the chi-square statistic \(\chi^2\). This is illustrated in  Figure 1 corresponding to the  1$\sigma$ and 2$\sigma$ error bars obtained from the $CC$ data. 	
	
\begin{figure}
	\centering
	\includegraphics[width=100 mm]{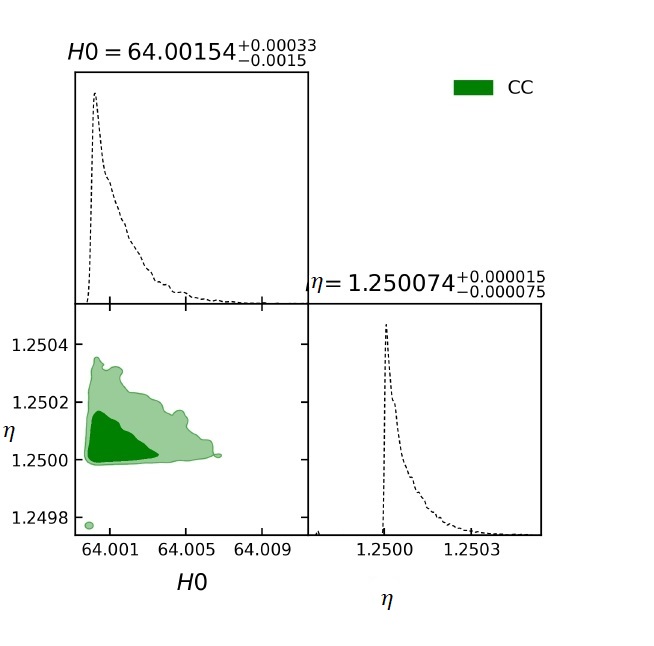}
	\caption{This figure corresponds to 1$\sigma$ and 2$\sigma$ error bars obtained from $CC$ data.} \label{fig_CC}
\end{figure}

	\subsection{Pantheon Dataset and Observational Constraints}
	The Pantheon dataset is a comprehensive compilation of 1048 Type Ia supernovae (SNe Ia) observations over a redshift range of \(0.01 < z < 2.3\). It is widely used to constrain cosmological parameters and test various cosmological models, including the standard \(\Lambda\)CDM model and models in  modified theories of gravity.
	The key observable in the Pantheon dataset is the \textit{distance modulus} \(\mu(z)\), which is related to the luminosity distance \(d_L(z)\) by:
	\begin{equation}
		\mu(z) = 5 \log_{10}\left[\frac{d_L(z)}{1 Mpc}\right] + 25,
	\end{equation}
	where \(d_L(z)\) is the luminosity distance in megaparsecs (Mpc). The luminosity distance depends on the cosmological model and is given by:
	\begin{equation}
		d_L(z) = (1 + z) \int_0^z \frac{\, dz'}{H(z')}
	\end{equation}	
where $z'$ is an arbitrary variable of integration and \(H(z')\) is the Hubble parameter.

Again we minimize the chi-square statistic to obtain the best-fit values of the model parameters:
\begin{equation}
	\chi_{SNIa}^2 = \sum_{i=1}^{N} \frac{[\mu_{\text{obs},i} - \mu_{\text{th},i}]^2}{\sigma_i^2},
\end{equation}
The best-fit parameters (\(H_0\), \(n\)) are those that minimize \(\chi^2\), yielding the maximum likelihood estimate of the parameters. In Figure 2, the likelihood contours at 1$\sigma$ and 2$\sigma$ levels for the Pantheon data are illustrated. We have illustrated only the Pantheon data, and not the $CC+SNIa$ data, as the values for this, for some reason or the other, were very close to those of the $CC$ dataset.

\begin{figure}
	\centering
	\includegraphics[width=100 mm]{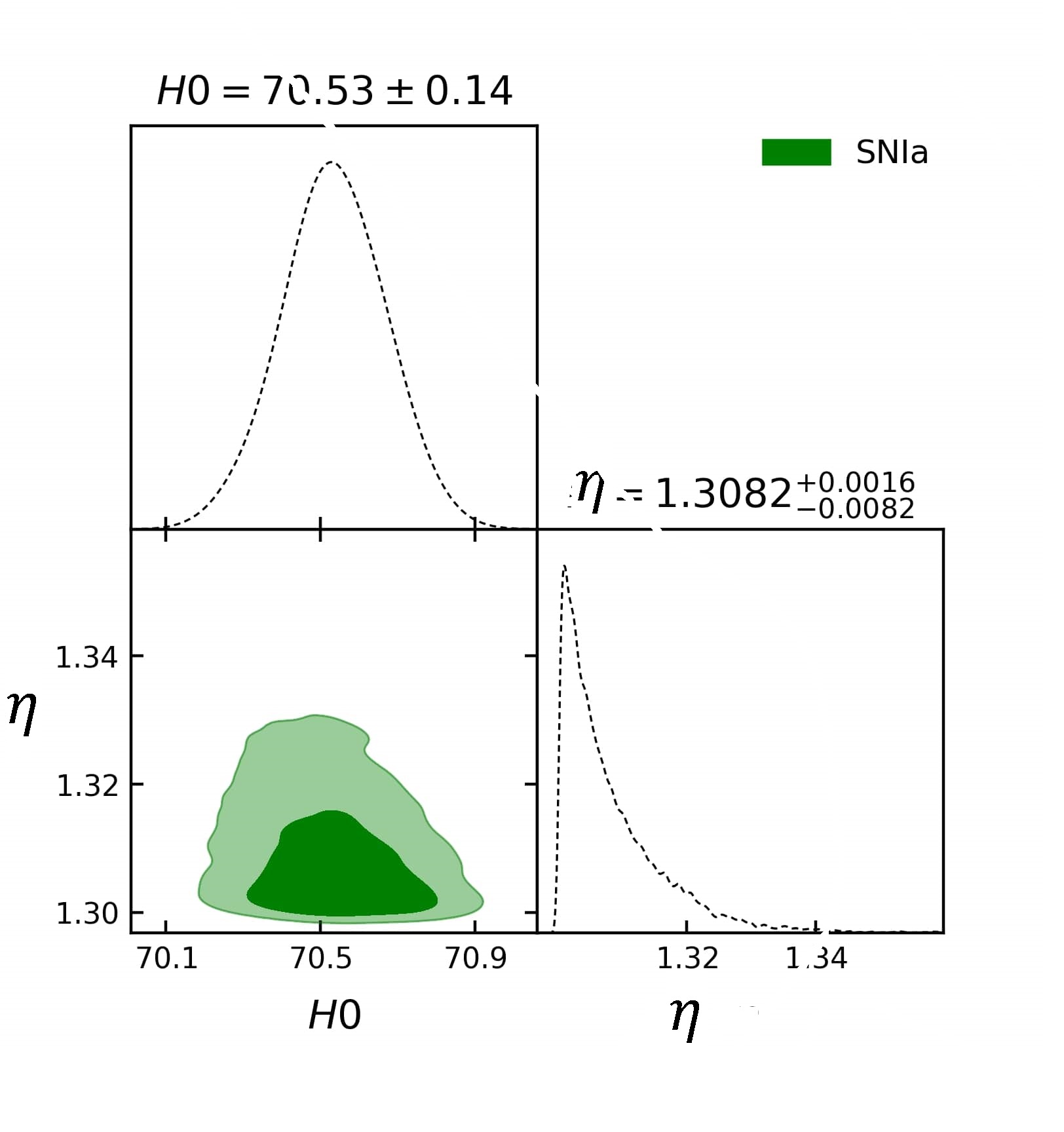}
	\caption{This figure corresponds to 1$\sigma$ and 2$\sigma$ error bars obtained from $SNIa$ datasets.} \label{fig_SN}
\end{figure}

\subsection{BAO Datasets}
Baryon Acoustic Oscillations (BAOs) are pressure waves produced by cosmological perturbations in the baryon-photon plasma during the recombination epoch, appearing as distinct peaks on large angular scales. In this study, we utilize the 26  BAO dataset points from the Six Degree Field Galaxy Survey (6dFGS), the Sloan Digital Sky Survey (SDSS), and the LOWZ samples of the Baryon Oscillation Spectroscopic Survey (BOSS) \cite{53}. The
surveys have provided highly accurate measurements of the positions of the BAO peaks in galaxy clustering at different redshifts.
The dilation scale, \(D_v(z)\) is given by
\begin{equation}
	D_v(z) = \left[ \frac{d_A^2(z) , cz}{H(z)} \right]^{1/3},
\end{equation}
The chi-square for BAO data,  \(\chi_{BAO}^2\)  , is expressed as
\begin{equation}
	\chi^2_{\text{BAO}} = \mathbf{X}^\mathrm{T} \mathbf{C}^{-1} \mathbf{X},
\end{equation}
where the vector $X$ is the ratio of the luminosity distance $d_A(z)$ to the dilaton scale  $D_V$, and  \(C\) denotes the covariance matrix \cite{55}.

\subsection{ Hz+SNeIa+BAO datasets}
In addition, use is made of \(\chi^2_{total}\) in order to get, from the combined $H(z)$, Pantheon+ and   $BAO$ data,  constraints on the parameters $H_0$, and  $\eta$. So, the definition of the required chi-square function is: 
\begin{equation}
	\chi^2_{total}= \chi^2_{Hz} + \chi^2_{SNe}+\chi^2_{BAO}     
\end{equation}

\begin{figure}
	\centering
	\includegraphics[width=100 mm]{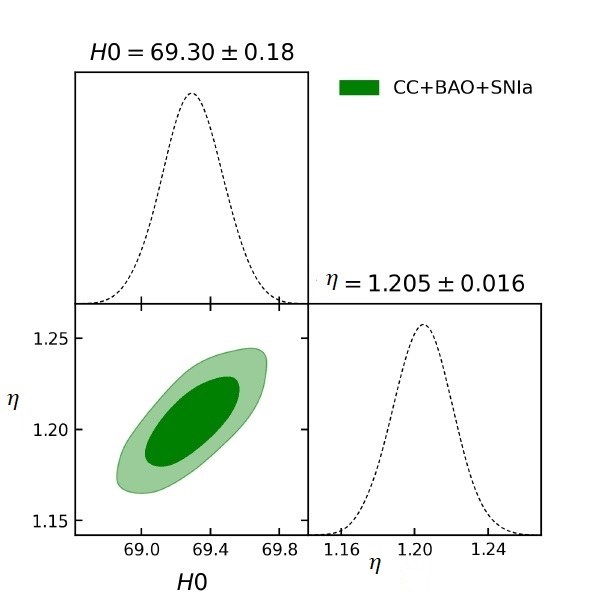}
	\caption{This figure corresponds to 1$\sigma$ and 2$\sigma$ error bars obtained from $CC+SNIA+BAO$ data.} \label{fig_CC+SNIA+BAO}
\end{figure}

Gaussian priors are used as follows for analysing: \([1,5]\) for \(\eta\),  and  \([60,80]\) for \(H_0\). For the $CC$ data, $SNIa$  and $CC+SNIa+BAO$ data, the corresponding $1-\sigma$ and $2-\sigma$ contour plots,  are shown  in Figures \ref{fig_CC}, \ref{fig_SN} and \ref{fig_CC+SNIA+BAO}, respectively.

With  \(68\%\) confidence limit, the results are as follows: \(H_0=64.00\) and \(\eta=1.25\), for the \(CC\) data; \(H_0=70.54\pm 0.14\) and \(\eta=1.3082^{+0.0016}_{-0.0081}\) for the  SNIa data; \(H_0=69.30\pm0.02\) and \(\eta=1.20\pm0.02\),  for the CC+SNIa+BAO data. We note that the values that Pawde et al \cite{Pawde} have used for $\eta$ are 1.4, 1.6 and 1.8, which are somewhat outside the values as obtained from observations. 

In Figure 4, we have displayed the $H(z)$ curve using the $CC+SNIa+BAO$ data, which shows the close correspondence between the $\Lambda$CDM model and our data. Figure 5 shows the $\mu(z)$ curve using the combined $CC+SNIa+BAO$ data, and once again we see a  close relationship with the $\Lambda$CDM model.

\begin{figure}
	\centering
	\includegraphics[width=100 mm]{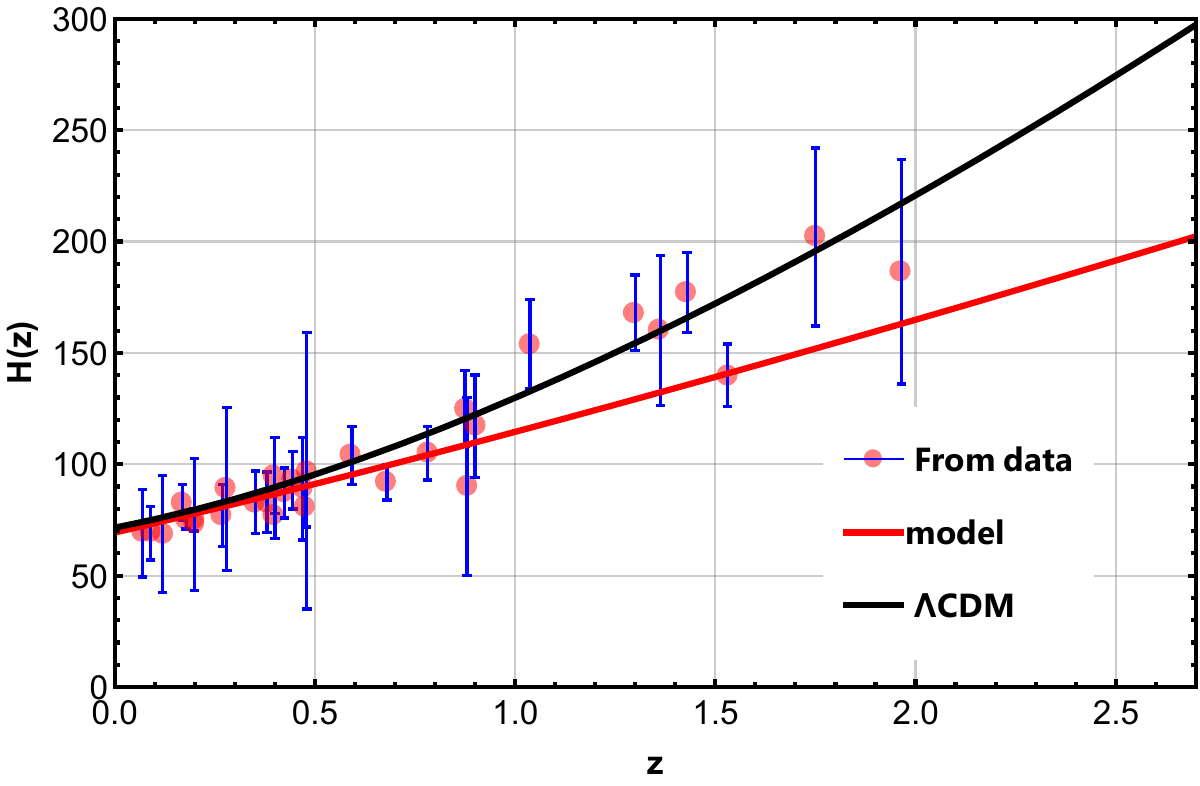}
	\caption{This figure displays the $H(z)$ curve using CC+SNIa+BAO data.} \label{fig_H}
\end{figure}
\begin{figure}
	\centering
	\includegraphics[width=100 mm]{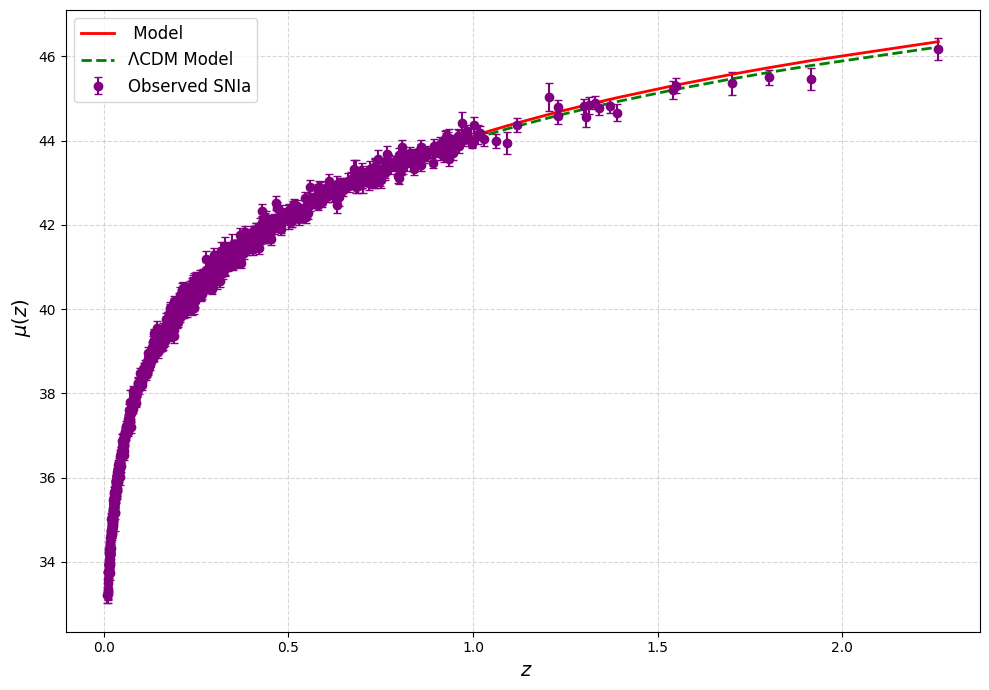}
	\caption{This figure displays the $\mu(z)$ curve using CC+SNIa+BAO data.} \label{fig_S}
\end{figure}

\section{Cosmographic parameters}
Cosmography is an important branch of cosmology for which a description of the universe is sought independently of any cosmological model \cite{Visser,Bamba,Capozziello,Dunsby,Vita,Bolotin,Hu}, and references therein. 
One expands a suitable parameter, such as the equation of state parameter, deceleration parameter, Hubble parameter, or scale factor,   in  a Taylor series around the current time. This enables comparing these quantities with observational data.  So,  cosmography is a technique which, in a certain sense, is independent of any specific cosmological model. It may be utilised to match parameters with observational constraints. However, there are some limitations on its use  as far as re-constructions of models are concerned. The easiest parameter to use for expansion in a Taylor series is the scale factor. The terms of the Taylor series can be compared to various cosmographic parameters as follows:

\begin{subequations}\label{CSdef}
	\begin{align}
		H \equiv \frac{1}{a} \frac{d a}{d t}\,,\quad q \equiv -\frac{1}{a H^2} \frac{d^2a}{d t^2}\,,\quad j  \equiv r  \equiv \frac{1}{a H^3} \frac{d^3a}{d t^3}\,,\\
		s \equiv \frac{1}{a H^4} \frac{d^4a}{d t^4}\,,\quad l  \equiv \frac{1}{a H^5} \frac{d^5a}{d t^5}\,,\quad m \equiv \frac{1}{a H^6} \frac{d^6a}{d t^6}\,~ etc.
	\end{align}
\end{subequations}
where  $j$ (or also called $r$) is the jerk parameter (or jolt), $s$ the snap (or facetiously called jounce), $l$ the lerk (or facetiously called crackle) and $m$ the pop parameter.  Less common alternative names for the  $j$ parameter    are pulse, impulse, bounce, surge, shock and super-acceleration.

We now plot several of these parameters using $\eta=1.20$ from the combined data, and compare with the $\Lambda$CDM model. For all subsequant Figures, we use this value for $\eta = 1.20$. The other values for $\eta$ do not make much of a difference to the picture.
\begin{figure}
	\centering
	\includegraphics[width=100 mm]{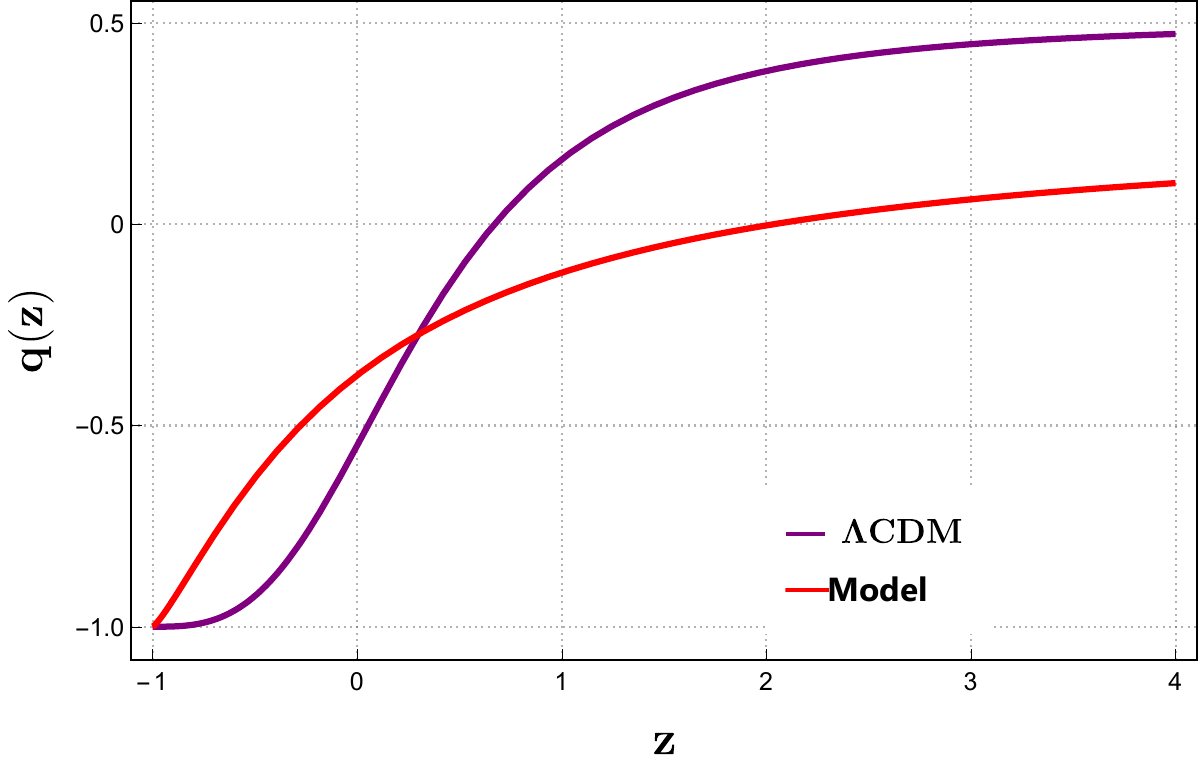}
	\caption{Deceleration parameter q vs redshift z} \label{fig_q}
\end{figure}

Firstly, in Figure 6, we plot the deceleration parameter against redshift using the combined data value $\eta = 1.20$.  From equation (\ref{qzformeq}), we see that the current value of the deceleration parameter $q$ is  $q(0) = -0.38$, using the value $\eta = 1.25$ (from the $CC$  data). 
The transition redshift $z_t$, i.e., the redshift $z_t$ for which the universe changes from deceleration to acceleration, is given by $q(z_t) = 0$, from which we get $z_t = 2.03$. For the SNIa dataset, $\eta = 1.31$, we get $q(0) = -0.345$, $z_t = 1.44. $ For $\eta = 1.20$ from the CC+SNIa+BAO data, we get $q(0) = -0.40$ and $z_t = 2.82$. 
The $\Lambda$CDM model has $q(0) \sim -0.55, $ $z_t = \left[ \frac{2(1-\Omega_{\Lambda 0}}{\Omega_{m0}}\right]^{(1/3)}-1 = 0.67$ using the Planck 2018 data \cite{Aghanim}. The values obtained with the present choice of deceleration prameter  $q$ in this model are higher than the $\Lambda$CDM model for $q(0)$, and much higher for the transition redshift.  $z_t$. They do not match, indicating that the present model is not viable. This is a very serious problem with the model of Pawde \cite{Pawde}.

	The jerk parameter is given by 
	\begin{equation}
		j=1+\frac{\eta^2}{\left( \frac{1}{1+z}\right)^\eta +1}+\frac{\eta^2}{\left(\left( \frac{1}{1+z}\right)^\eta +1\right)^2}-\frac{3\eta}{\left( \frac{1}{1+z}\right)^\eta +1}
	\end{equation}

 The snap parameter is given by
	\begin{equation}
		s= \frac{2\eta^2 \left( \left( \frac{1}{1+z}\right)^\eta +1\right)-6\eta}{6\eta -9\left(\left( \frac{1}{1+z}\right)^\eta +1\right)}
	\end{equation}
	
\begin{figure}
	\centering
	\includegraphics[width=100 mm,height=60 mm]{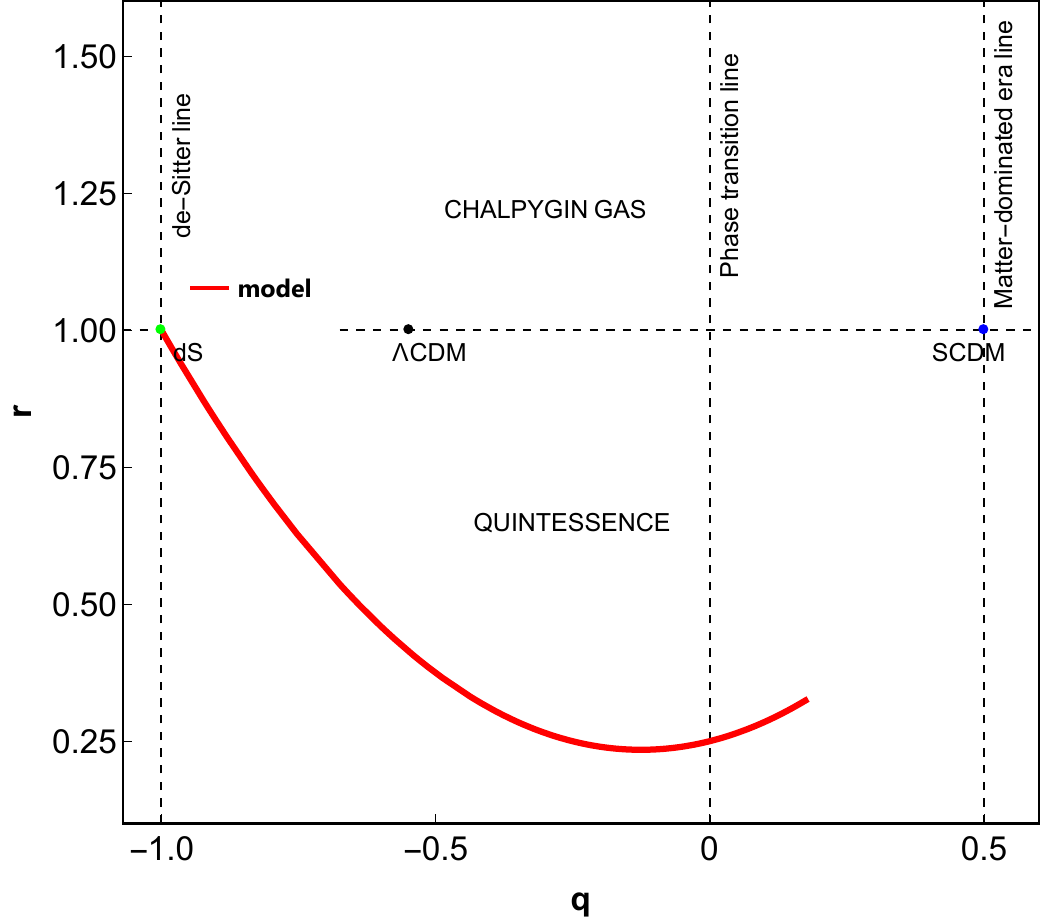}
	\caption{State-finder parameter r against deceleration parameter $q$} \label{fig_12rq}
\end{figure}

The evolution of the $r-q$ curve (Note that the formula for $r$ is the same as that of $j$) is given in Figure 7.  The curve does not continue just after $q>0$. This is due to a singularity at that point, and this is also reflected in the corresponding Figure of Pawde et al.  \cite{Pawde}. For calrity we point out the in Figures 7 and 8, the region above the line $r=1$ is the Chaplygin gas region, and the region below that line is the quintessence region.

\begin{figure}
	\centering
	\includegraphics[width=100 mm,height=60 mm]{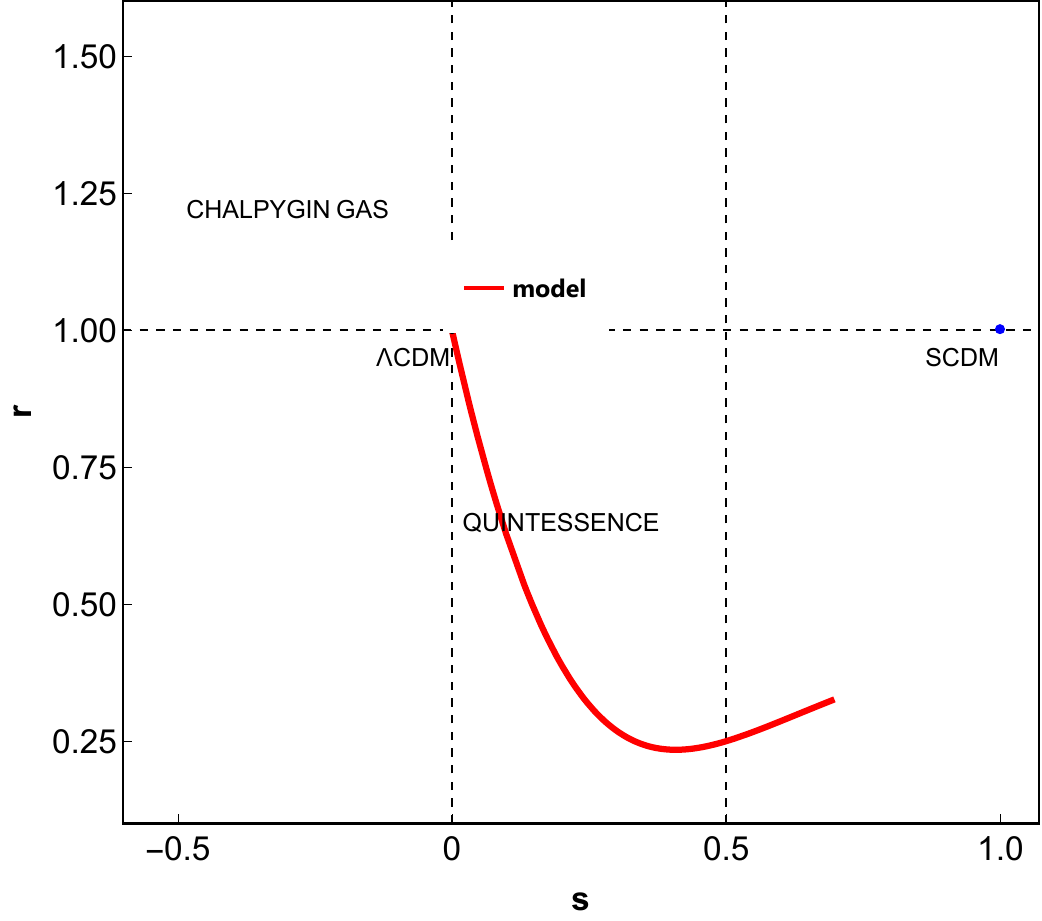}
	\caption{State-finder parameter $(r,s)$ curve}  \label{fig_12r}
\end{figure}

In Figure 8 above, we have illustrated the $(r,s)$ state-finder parameter curve. In Figures 6-8, we used the CC+SNIa+BAO data to plot the curves.	
	
 \section{Physical parameters}
	
	To plot the energy density $\rho$ as a function of $z$, we make use of Eq. (\ref{rhoz}):
	\begin{equation}
		\rho(z)=\left[ \frac{9\left( \frac{1}{1+z}\right)^{-2\eta}\left( \left( \frac{1}{1+z}\right)^{\eta}+1\right)^2}{4(1-2\alpha)}-\beta\right]^{\frac{1}{\alpha}}
	\end{equation}
	\begin{figure}
		\centering
		\includegraphics[width=100 mm,height=60 mm]{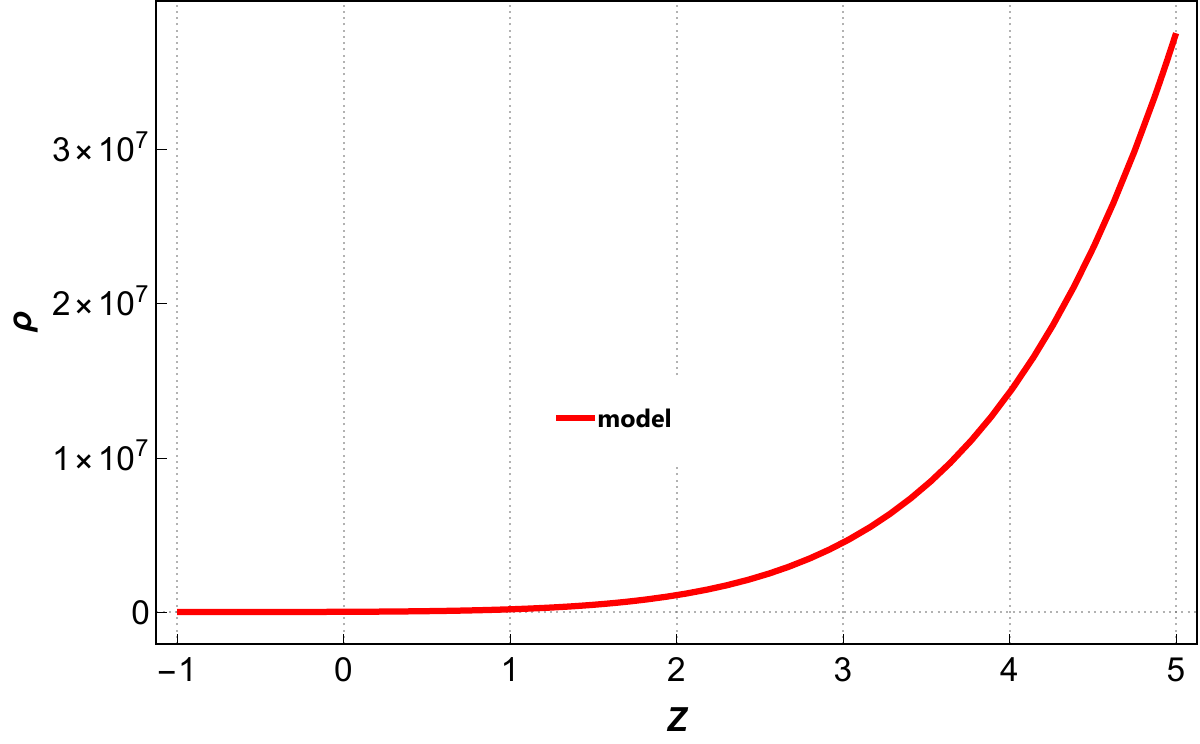}
		\caption{Energy density vs redshift}  \label{fig_rho}
	\end{figure}
	
	In Figure 9, we have plotted the energy density $\rho(z)$  from Eq. (\ref{rhoz}) using $\eta = 1.20$ (from the $CC+SNIa+BAO$ data). We have taken $\alpha = 0.42, ~ \beta = 0.6$ \cite{Pawar} to facilitate comparison with the paper \cite{Pawde}.  We use this value for all subsequent figures as well). The energy density remains positive throughout the evolution of the universe. It is monotonically decreasing, and tends to zero at  in the future ($z \rightarrow -1$). This is typical behaviour of a continuously expanding model in which the matter content of the universe gets diluted continuously as the universe expands.  Indeed, in Figure 10, the pressure is plotted against redshift from Eq.  (\ref{presz}):	

\begin{equation}
	p(z) =\frac{\left[
		\frac{\alpha \beta}{1 - 2\alpha} 
		-  \frac{9\alpha\left( \frac{1}{1+z}\right)^{-2\eta}\left( \left( \frac{1}{1+z}\right)^{\eta}+1\right)^2}{4(1-2\alpha)}
		- 4\eta\left( \frac{1}{1+z}\right)^{-2\eta}\left( \left( \frac{1}{1+z}\right)^{\eta}+1\right)
		\right]}{\alpha\left[ \frac{9\left( \frac{1}{1+z}\right)^{-2\eta}\left( \left( \frac{1}{1+z}\right)^{\eta}+1\right)^2}{4(1-2\alpha)}-\beta\right]^{\frac{\alpha -1}{\alpha}}} 
\end{equation}
It starts off with a fairly ``large" negative value, and then evolves monotonically towards smaller negative values, eventually approaching zero at present and in future. This kind of behaviour is typical of modified gravity, and does not require an explicit dark energy exotic fluid component to achieve late-time accelerated expansion. Rather, it is driven by geometric contributions.

\begin{figure}
	\centering
	\includegraphics[width=100 mm,height=60 mm]{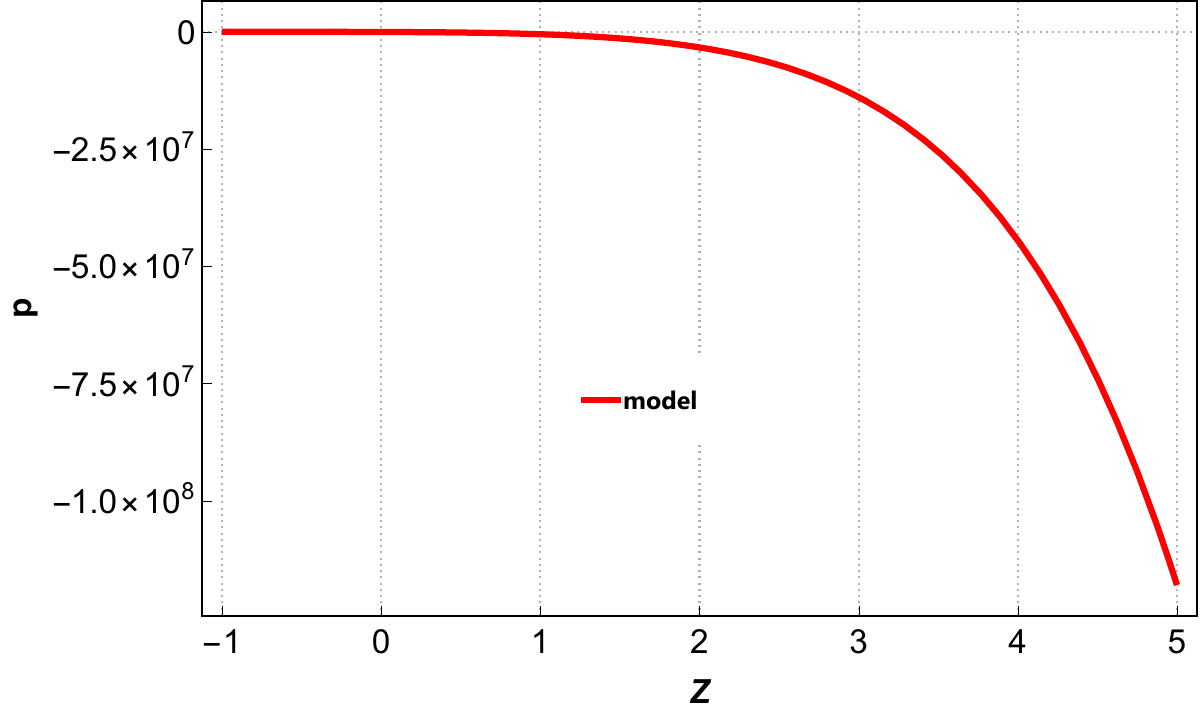}
	\caption{Pressure vs redshift}  \label{fig_p}
\end{figure}
\begin{figure}
	\centering
	\includegraphics[width=100 mm,height=60 mm]{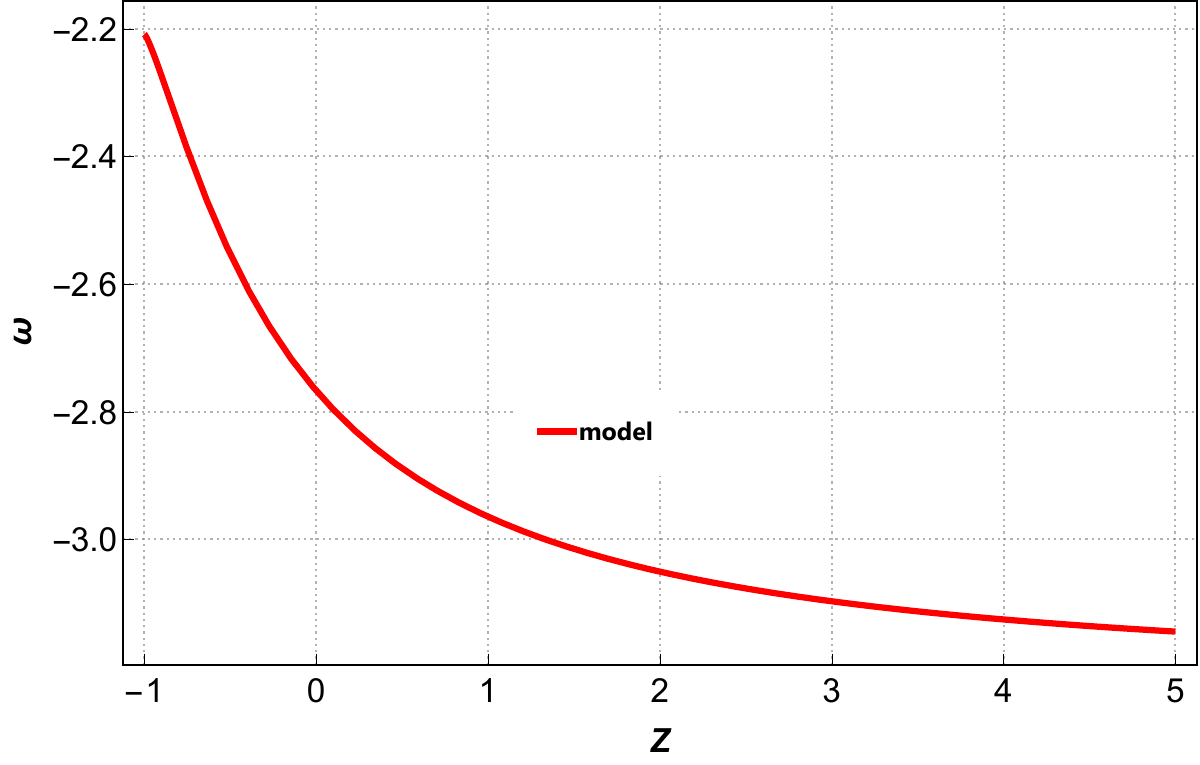}
	\caption{Equation of state parameter vs redshift}  \label{fig_w}
\end{figure}

The equation of state parameter $\omega = p/\rho$  plays a crucial role as it relates to the
behavior of different components of the universe. such as radiation, dark matter and dark energy.  The EoS parameter is not, in general, constant, but can vary depending on the matter content. For radiation, the EoS is 1/3, for ordinary matter (baryonic and non-relativistic matter), it is approximately w = 0, and for dark energy it is normally -1, as in the $\Lambda$CDM model. This is associated with the cosmological constant $\Lambda$, with an exotic fluid with constant energy density, and negative pressure. There are a further two types of dark energy associated with particular values of $\omega$.	

\begin{itemize}
	\item The one is phantom energy, for which \(\omega <-1\), which 
	represents a form of dark energy with even more negative pressure than the cosmological constant.  If phantom energy keeps dominating at late times, it may result in a catastrophic situation known as the Big Rip, wherein the accelerated expansion becomes so extreme that it tears galaxies, stars, planets, and eventually even atomic structures apart. This unbounded expansion contrasts with more stable scenarios predicted by dark energy with \(\omega \geq -1\). 
	\item The other category of dark energy is quintessence, usually with an equation of state in the range $-1 < \omega < -1/3$. Quintessence can be similar to the cosmological constant, but can also give rise to different behaviour. 
\end{itemize}
In Figure 11, we have sketched the equation of state parameter $\omega(z)$ using Eq. (\ref{eosz}). It is seen that it starts off with a large negative value,  increases with time, and approaches the $\Lambda$CDM value of -1 in future. The current value of the equation of state, $\omega(0) = -2.78$, which is  less than the $\Lambda$CDM value of $\omega(0) = -1$. The equation of state falls into the category $\omega \leq -1$, i.e., phantom. For this category, as we indicated earlier, the future does not look bright. The energy density of phantom energy
increases as the universe expands, leading to a
``big rip" scenario where the universe is ultimately torn apart. The concept of phantom DE is highly speculative and is not supported by current observational data. 

It appears that in \cite{Pawde}, the authors have listed the equation for $\omega(z)$ as its inverse, and have also plotted the inverse.

\section{Energy conditions}
In GR,  one expects physically reasonable matter and energy distributions. For this, a set of  energy conditions, which are  a set of constraints on the stress-energy tensor $T_{ij}$ are imposed  to ensure physically reasonable matter. These
conditions help in understanding the behavior of matter and energy under various circumstances, especially in the context of gravitational collapse, black holes, and cosmological models. Here, the primary role of these energy conditions is to relate to the  the accelerated expansion of the universe. These conditions originate from
the well-established Raychaudhuri equation, which describe the behavior of geodesic congruences (families) in a given spacetime, illustrating the focusing of geodesic flows due to gravity. It must be remembered that these energy conditions are derived for GR, and for a perfect fluid, so that they may not necessarily hold for modified gravity theories, or for imperfect fluids. Nonetheless, it is instructive to examine these energy conditions. 

\subsection{Weak energy condition}
The \textbf{Weak Energy Condition (WEC)} ensures that the energy density as measured by any timelike observer is non-negative. Mathematically, the WEC states that for any timelike vector $v^\mu$, the energy-momentum tensor $T_{\mu \nu}$ satisfies:
\begin{equation}
	T_{\mu \nu} v^\mu v^\nu \geq 0.
\end{equation}

In cosmology, for a perfect fluid with energy density $\rho$ and pressure $p$, the energy-momentum tensor is given by:
\begin{equation}
	T_{\mu \nu} = (\rho + p) u_\mu u_\nu + p g_{\mu \nu},
\end{equation}
where $u^\mu$ is the four-velocity of the fluid and $g_{\mu \nu}$ is the metric tensor.
Substituting this into the WEC condition, we obtain:
\begin{equation}
	T_{\mu \nu} v^\mu v^\nu = (\rho + p)(u_\mu v^\mu)^2 + p(v_\mu v^\mu).
\end{equation}

For a timelike vector $v^\mu$ (where $v_\mu v^\mu < 0$), the WEC simplifies to:
\begin{equation}
	\rho + p \geq 0 \quad \text{and} \quad \rho \geq 0.
\end{equation}
The quantity $\rho + p$ in our model works out to be:
\[
\rho +p= \left[\left[ \frac{9\left( \frac{1}{1+z}\right)^{-2\eta}\left( \left( \frac{1}{1+z}\right)^{\eta}+1\right)^2}{4(1-2\alpha)}-\beta\right]^{\frac{1}{\alpha}}\right]
\]
 
$\\ +\left[\frac{\left[
	\frac{\alpha \beta}{1 - 2\alpha} 
	-  \frac{9\alpha\left( \frac{1}{1+z}\right)^{-2\eta}\left( \left( \frac{1}{1+z}\right)^{\eta}+1\right)^2}{4(1-2\alpha)}
	- 4\eta\left( \frac{1}{1+z}\right)^{-2\eta}\left( \left( \frac{1}{1+z}\right)^{\eta}+1\right)
	\right]}{\alpha\left[ \frac{9\left( \frac{1}{1+z}\right)^{-2\eta}\left( \left( \frac{1}{1+z}\right)^{\eta}+1\right)^2}{4(1-2\alpha)}-\beta\right]^{\frac{\alpha -1}{\alpha}}}\right]$

This energy condition has been plotted in Figure 12, from which it can be seen that it is violated in our model.

\begin{figure}
	\centering
	\includegraphics[width=100 mm,height=60 mm]{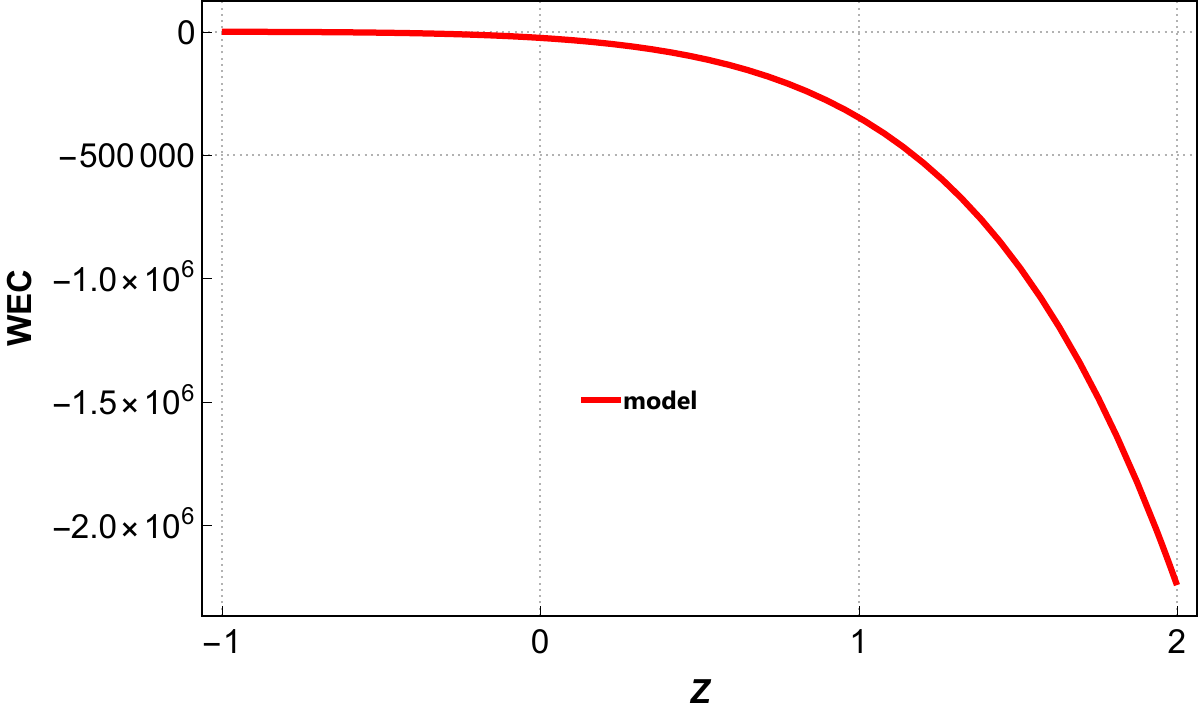}
	\caption{WEC vs redshift}  \label{fig_wec}
\end{figure}
When the  WEC is not satisfied in general relativistic cosmology, it means that the energy-momentum tensor \(T_{\mu\nu}\) does not satisfy the usual expectations for physically realistic matter and energy. The WEC demands that the energy density \(\rho\) be non-negative and that the pressure \(p\) be greater than or equal to the negative of the energy density:\(p\geq -\rho\). In GR, violation of the WEC implies that there must exist some forms of matter or energy which is exotic, including negative  pressure. This may lead to behavior which is unphysical in nature, including the possibility of phantom energy which as we have mentioned previously leads to a big rip singularity. The breakdown of the WEC would also mean the existence of closed timelike curves or other paradoxical phenomena in the spacetime.

\subsection{Dominant Energy condition}
The \textbf{Dominant Energy Condition (DEC)} is another fundamental energy condition in general relativistic cosmology. It requires that energy density is non-negative and that energy flows in a causal manner, i.e., it cannot propagate faster than the speed of light. Mathematically, the DEC states that for any timelike or null vector $v^\mu$, the energy-momentum tensor $T_{\mu \nu}$ satisfies:
\begin{equation}
	T_{\mu \nu} v^\mu v^\nu \geq 0,
\end{equation}
and the energy flux vector $T^\mu_{\ \nu} v^\nu$ is non-spacelike:
\begin{equation}
	T^\mu_{\ \nu} v^\nu \quad \text{is timelike or null}.
\end{equation}
Fronm the above, for the DEC to hold, the following conditions must be satisfied. 
The energy density must be non-negative:
\begin{equation}
	\rho \geq 0.
\end{equation}
and the 
energy flux must be causal, leading to the condition:
\begin{equation}
	\rho \geq |p|.
\end{equation}
This inequality implies two sub-conditions:
\begin{equation}
	\rho - p \geq 0 \quad \text{(non-negative radial energy density)},
\end{equation}

\begin{equation}
	\rho + p \geq 0 \quad \text{(non-negative tangential energy density)}.
\end{equation}

For our model
\[
\rho -p= \left[\left[ \frac{9\left( \frac{1}{1+z}\right)^{-2\eta}\left( \left( \frac{1}{1+z}\right)^{\eta}+1\right)^2}{4(1-2\alpha)}-\beta\right]^{\frac{1}{\alpha}}\right]
\]

 \[
\\ - \left[\frac{\left[
	\frac{\alpha \beta}{1 - 2\alpha} 
	-  \frac{9\alpha\left( \frac{1}{1+z}\right)^{-2\eta}\left( \left( \frac{1}{1+z}\right)^{\eta}+1\right)^2}{4(1-2\alpha)}
	- 4\eta\left( \frac{1}{1+z}\right)^{-2\eta}\left( \left( \frac{1}{1+z}\right)^{\eta}+1\right)
	\right]}{\alpha\left[ \frac{9\left( \frac{1}{1+z}\right)^{-2\eta}\left( \left( \frac{1}{1+z}\right)^{\eta}+1\right)^2}{4(1-2\alpha)}-\beta\right]^{\frac{\alpha -1}{\alpha}}}  \right]
\]

and this is sketched in Figure 13, from which is may be seen that the DEC is satisfied for our model.
\begin{figure}
	\centering
	\includegraphics[width=100 mm,height=60 mm]{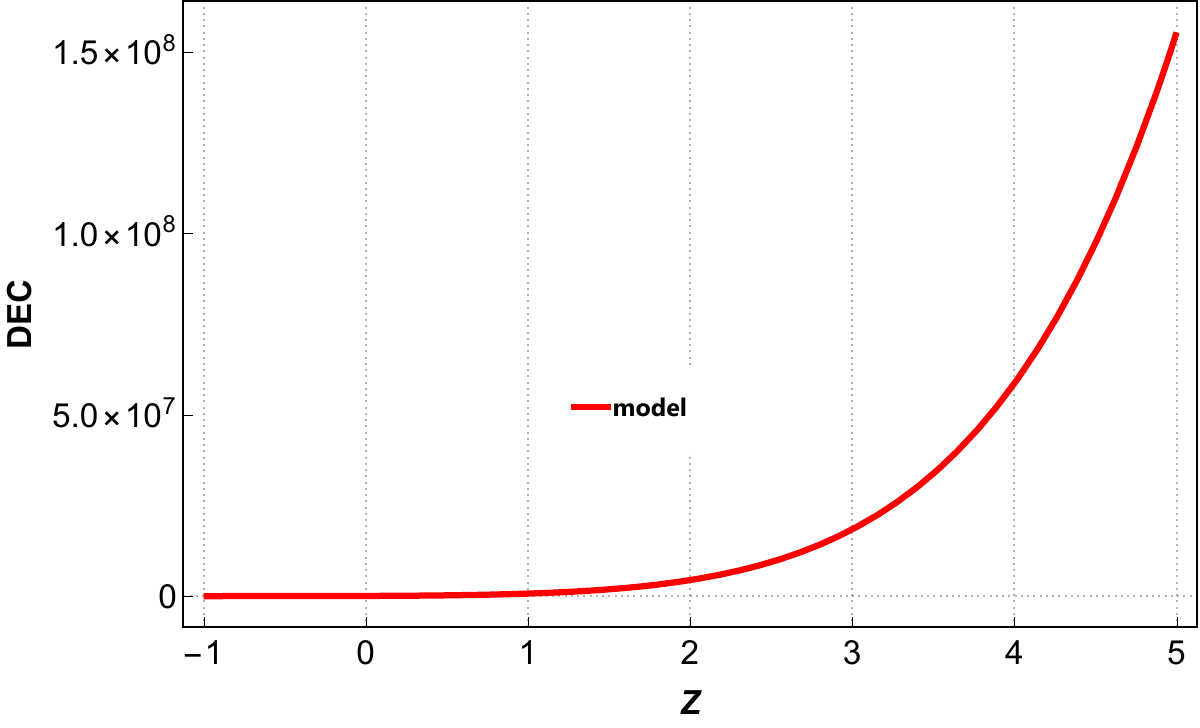}
	\caption{DEC vs redshift}  \label{fig_dec}
\end{figure}

 \subsection{Strong Energy condition}
The \textbf{Strong Energy Condition (SEC)} places constraints on the curvature of spacetime produced by matter and energy. Unlike other energy conditions, the SEC does not require the energy density to be positive but imposes conditions on the combination of energy density and pressure, which directly affect gravitational attraction. 
The SEC states that for any timelike vector $v^\mu$, the Ricci tensor $R_{\mu \nu}$ satisfies:
\begin{equation}
	R_{\mu \nu} v^\mu v^\nu \geq 0.
\end{equation}
Using the Einstein field equations:
\begin{equation}
	R_{\mu \nu} - \frac{1}{2} R g_{\mu \nu} = 8 \pi G T_{\mu \nu},
\end{equation}
where $R$ is the Ricci scalar and $T_{\mu \nu}$ is the energy-momentum tensor, the SEC can be expressed in terms of the energy-momentum tensor as:
\begin{equation}
	\left( T_{\mu \nu} - \frac{1}{2} T g_{\mu \nu} \right) v^\mu v^\nu \geq 0,
\end{equation}
where $T = g^{\mu \nu} T_{\mu \nu}$ is the trace of the energy-momentum tensor.

For a perfect fluid, 
\begin{equation}
	\left[(\rho + p) u_\mu u_\nu + p g_{\mu \nu} - \frac{1}{2} (\rho - 3p) g_{\mu \nu}\right] v^\mu v^\nu \geq 0.
\end{equation}
For a timelike vector $v^\mu = u^\mu$, this simplifies to:
\begin{equation}
	\rho + 3p \geq 0.
\end{equation}

For our model:
\[
\rho +3p= \left[\left[ \frac{9\left( \frac{1}{1+z}\right)^{-2\eta}\left( \left( \frac{1}{1+z}\right)^{\eta}+1\right)^2}{4(1-2\alpha)}-\beta\right]^{\frac{1}{\alpha}}\right]
\]
\[
\\+3\left[\frac{\left[
	\frac{\alpha \beta}{1 - 2\alpha} 
	-  \frac{9\alpha\left( \frac{1}{1+z}\right)^{-2\eta}\left( \left( \frac{1}{1+z}\right)^{\eta}+1\right)^2}{4(1-2\alpha)}
	- 4\eta\left( \frac{1}{1+z}\right)^{-2\eta}\left( \left( \frac{1}{1+z}\right)^{\eta}+1\right)
	\right]}{\alpha\left[ \frac{9\left( \frac{1}{1+z}\right)^{-2\eta}\left( \left( \frac{1}{1+z}\right)^{\eta}+1\right)^2}{4(1-2\alpha)}-\beta\right]^{\frac{\alpha -1}{\alpha}}}  \right]
\]

and we have plotted this in Figure 14.
\begin{figure}
	\centering
	\includegraphics[width=100 mm,height=60 mm]{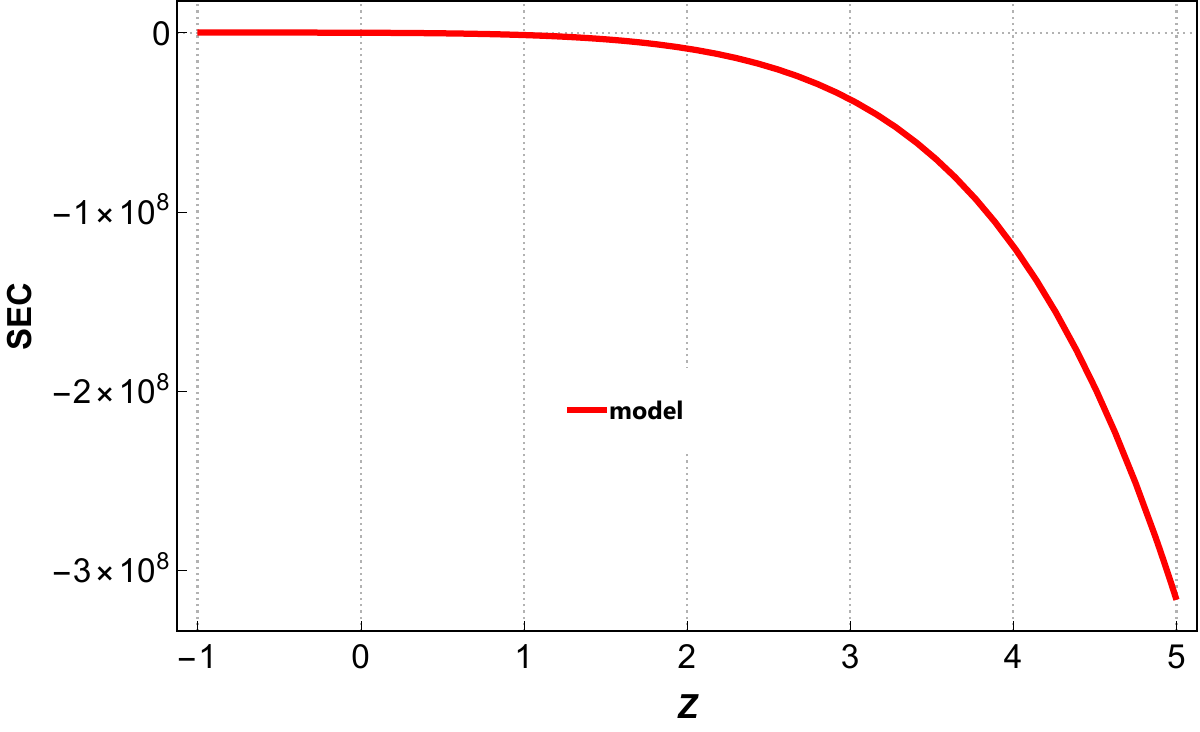}
	\caption{SEC vs redshift}  \label{fig_sec}
\end{figure}
The SEC in this model is violated. In general relativistic cosmology, when the SEC is not satisfied, then it denotes the existence of some exotic kinds of energy or matter, which do not behave according to the expected behavior by ordinary matter. The SEC  guarantees that gravity is attractive and the expansion or contraction of the universe behaves in a physically consistent manner. A violation of the SEC leads to  scenarios, such as repulsive gravity or the possibility of exotic matter with negative  pressures. In the $\Lambda$CDM model, a violation of the SEC occurs at late times, associated with the cosmological constant and the current acceleration of the universe. In modified gravity theories, a violation of the SEC can occur for all time, like in our case, but there is still the early deceleration.

\section{Discussion}

In this work, we have applied observational constraints to the model discussed by Pawde et al \cite{Pawde} in $f(R,L_m)$ gravity. These authors chose a particular form for the deceleration parameter of the form Eq. (\ref{defnqa}). Those authors did not apply observational constraints to determine the parameter $\eta$ of the model, but rather chose values of $\eta$ equal to 1.4, 1.6 and 1.8, respectively. We have applied observational constraints from CC, SNIa, and BAO datasets to determine $\eta$. We find that $\eta =1.25$ for the CC dataset, $1.31$ for the SNIa dataset, and $1.2$ for the combined CC+SNIa+BAO datasets.   The  values  of \cite{Pawde} are somewhat higher than the above-mentioned. The current value for the deceleration parameter, as well as the transition redshift do not match with those of the $\lambda$CDM model, and this presents a serious problem for the model studied here.
	
	We began by giving some background in the introduction to the particular form of $q$. Then in Section 2, we discussed the basic formalism of $f(R,L_m)$ gravity. Then in section 3, we studied the LRS Bianchi I metric in the theory, giving the field equations. Then we choose the required form for $f(R,L_m)$ gravity in section 4. In section 5, we gave the forms for the  scale factor $a$, Hubble parameter $H$ and $q$ both in terms of the time $t$ and redshift $z$. A detailed motivation for the chosen parametrization of $q(z)$ was given. Also the equations for the physical parameters energy density $\rho$, pressure $p$ and equation of state paramerter $\omega$ were given.

The next section deals with the observational constraints that we find using the CC, SNIa and the combined CC+SNIa+BAO datasets. Best fit values for the present Hubble parameter $H0$ and the parameter $\eta$ were determined. We then plotted the $H(z)$ and $\mu(z)$ graphs in Figures 4 and 5, respectively.  For subsequent analysis, we chose the value $\eta = 1.20$ from the combined set (the other values do not affect the picture to any large extent).  We have plotted curves of  the relevant cosmographic parameters $(q, (r,Q), and (r,s))$ in Figures 6-8. These Figures show how the model differs from the $\Lambda$CDM model, and also gives some idea of how the model evolves. Then, we discuss and plot the physical parameters, the energy density, pressure and equation of state. The energy density is a positive monotonically decreasing function, whereas the pressure is a monotonically increasing negative function. This is in contrast to the $\Lambda$CDM model, for which the pressure shows a change from positive to negative, signifying the emergence and evolution of dark energy. The equation of state parameter is also an increasing function of time, and is negative throughout. The present value of the equation of state parameter $\omega_0 = -2.78$, is also not in keeping with the $\Lambda$CDM model ($\omega_0 = -1$ and observations: $\omega_0= -1.03 \pm 0.03$ \cite{Aghanim}. The equation of state parameter indicates that the model falls into the category of phantom dark energy, which is not well supported by observations. 

In conclusion, we can say that the present model and choice of deceleration parameter $q$ does not fit in with observational constraints. We are currently attempting to determine how to modify the choice of  $q$ so that a better model could be obtained. Most likely, the reason for this could be that there is only one parameter in the model (apart from the current Hubble parameter), and perhaps one needs more than one parameter.

\end{document}